\documentstyle[12pt,epsf]{article}
\begin{document}
\baselineskip 0.8cm

\newcommand{\gsim}{ \mathop{}_{\textstyle \sim}^{\textstyle >} }
\newcommand{\lsim}{ \mathop{}_{\textstyle \sim}^{\textstyle <} }
\newcommand{\vev}[1]{ \langle {#1} \rangle }
\newcommand{\EV}{ {\rm eV} }
\newcommand{\KEV}{ {\rm keV} }
\newcommand{\MEV}{ {\rm MeV} }
\newcommand{\GEV}{ {\rm GeV} }
\newcommand{\TEV}{ {\rm TeV} }
\def\tr{\mathop{\rm tr}\nolimits}
\def\Tr{\mathop{\rm Tr}\nolimits}
\def\Re{\mathop{\rm Re}\nolimits}
\def\Im{\mathop{\rm Im}\nolimits}
\setcounter{footnote}{1}

\begin{titlepage}

\begin{flushright}
UT-895\\
\end{flushright}

\vskip 2cm
\begin{center}
{\large \bf  Bulk U(1) Messenger}
\vskip 1.2cm
Yasunori Nomura$^1$ and T.~Yanagida$^{1,2}$

\vskip 0.4cm

$^{1}$ {\it Department of Physics, University of Tokyo, \\
         Tokyo 113-0033, Japan}\\
$^{2}$ {\it Research Center for the Early Universe, University of Tokyo,\\
         Tokyo 113-0033, Japan}
\vskip 1.5cm

\abstract{
 We propose a new U(1) gauge interaction in the bulk in higher
 dimensional spacetime, which transmits supersymmetry-breaking effects
 on the hidden brane to the observable our brane.
 We find that rather small gauge coupling constant of U(1)$_{\rm bulk}$, 
 $\alpha_{\rm bulk} \simeq 5 \times 10^{-4}$, is required for a
 successful phenomenology.
 This result implies the compactification length $L$ of the extra
 dimension to be $L^{-1} \simeq 2 \times 10^{15}~\GEV$ for
 $(4+1)$-dimensional spacetime.
 This large compactification length $L$ is a crucial ingredient to
 suppress unwanted flavor-changing neutral currents and hence our
 proposal is very consistent with the Randall-Sundrum brane-world
 scenario.
}

\end{center}
\end{titlepage}

\renewcommand{\thefootnote}{\arabic{footnote}}
\setcounter{footnote}{0}

%
%
%
%

Most of supersymmetry (SUSY) breaking models in supergravity (even
including gauge mediation models) assume a separation of the
SUSY-breaking and the SUSY standard-model sectors \cite{SUGRA, GMSB}.
However, the origin of the separation is not well understood, although
such a separation is crucial to obtain phenomenologically consistent
spectra for SUSY particles.

The brane world proposed by Randall and Sundrum \cite{RS} provides a
beautiful geometric explanation for the separation.
That is, the hidden and observable sectors live on different
three-dimensional branes separated by a gravitational bulk \cite{bulk}
in higher dimensional spacetime.
It has been, recently, claimed \cite{LS, RS} that this brane separation
produces the hidden and observable separation in the ``conformal'' frame
in supergravity, which was proposed long time ago from a
phenomenological ground \cite{IKYY}.

It is a crucial observation in Ref.~\cite{IKYY} that the above
separation in the ``conformal'' frame induces a no-scale type K\"ahler
potential\footnote{
The no-scale supergravity \cite{no-scale} adopts a specific form 
$f_H = Z + Z^{\dagger}$, where $Z$ is the superfield responsible for the 
SUSY breaking.
We assume the K\"ahler potential for the $Z$ field to be of the form 
$f_H = Z^{\dagger} Z + \cdots$, where the ellipsis denotes higher order
terms.}
in the Einstein frame,
\begin{eqnarray}
  K(\Phi_{\rm obs}, \Phi_{\rm obs}^{\dagger}, 
    \Phi_{\rm hid}, \Phi_{\rm hid}^{\dagger}) 
  = -3\log \left(1 
    - \frac{1}{3}f_O(\Phi_{\rm obs}, \Phi_{\rm obs}^{\dagger}) 
    - \frac{1}{3}f_H(\Phi_{\rm hid}, \Phi_{\rm hid}^{\dagger})\right).
\label{Kahler}
\end{eqnarray}
Here, $\Phi_{\rm obs}$ and $\Phi_{\rm hid}$ denote superfields in the
observable and hidden sectors, respectively.
With the above K\"ahler potential Eq.~(\ref{Kahler}) we easily show
\cite{IKYY} that all soft SUSY-breaking masses and $A$ terms in the
observable sector vanish in the limit of the zero cosmological constant.
All gaugino masses in the observable sector also vanish because of the
decoupling of the hidden superfield $Z$ from the gauge kinetic function
\cite{RS}.\footnote{
The gaugino-mediated SUSY breaking was proposed in Ref.~\cite{IKYY}.
See also recent works \cite{gaugino-med}.}

On the contrary, the SUSY-invariant $\mu$ term ($\mu H \bar{H}$)
naturally arises \cite{GM} from the K\"ahler potential if $f_O$ contains 
$f_O \supset H \bar{H}$, where $H$ and $\bar{H}$ are chiral superfields
of Higgs doublets.
This mechanism \cite{GM} produces the SUSY-invariant mass $\mu$ of the
order of the gravitino mass, {\it i.e.} $\mu \simeq m_{3/2}$.
This requires the gravitino mass $m_{3/2} \simeq 100~\GEV-1~\TEV$ for
the correct electroweak symmetry breaking.
With these gravitino masses the anomaly mediation \cite{RS, GLMR}
generates too small SUSY-breaking masses in the observable sector and
hence all SUSY particles in the observable sector except for the
Higgsinos and the gravitino remain almost massless.

In this letter we introduce a new U(1) gauge interaction in the bulk to
solve the above problem.\footnote{
A similar model in the brane world has been also discussed in
Ref.~\cite{INY}.}
The U(1)$_{\rm bulk}$ gauge superfield plays a role of messenger between
the hidden and the observable branes.
SUSY-breaking effects on the hidden brane are transmitted to the
observable brane through the bulk U(1)$_{\rm bulk}$ gauge interaction
and all of the gauginos, squarks and sleptons in the observable sector
acquire suitable SUSY-breaking masses.

It is well known that a similar U(1) gauge interaction is also used as a
messenger between the SUSY- breaking and observable sectors in a class
of gauge-mediated SUSY breaking models \cite{min-GMSB, NTY}.
Thus, it is quite natural to identify the above bulk U(1)$_{\rm bulk}$
gauge interaction with the messenger U(1)$_m$ gauge interaction in the
gauge mediation models.
We adopt a model proposed in Ref.~\cite{NTY}, and interpret it as a
low-energy effective theory of the brane world.
We then show that rather small gauge coupling 
$\alpha_{\rm bulk} \simeq 5 \times 10^{-4}$ of the U(1)$_{\rm bulk}$ is
required for a successful phenomenology.
This small coupling is regarded as a consequence of a large volume of
extra dimension.
In fact, the result implies the compactification length $L$ of the extra
dimension to be $L^{-1} \simeq 2 \times 10^{15}~\GEV$ for
$(4+1)$-dimensional spacetime and the fundamental scale $M_*$ is
determined as $M_* \simeq 2 \times 10^{17}~\GEV$ to reproduce the
gravitational scale 
$M_G \simeq (2 M_*^3 L)^{1/2} \simeq 2 \times 10^{18}~\GEV$.
Such a large compactification length is a crucial ingredient to suppress
the flavor-changing neutral currents (FCNC's) and hence our proposal is
very consistent with the brane-world scenario of Randall and Sundrum
\cite{RS}.

Let us first briefly review the gauge mediation model proposed in
Ref.~\cite{NTY}.
The model consists of three sectors: dynamical SUSY breaking (DSB)
sector, messenger sector, and the minimal SUSY standard model (MSSM)
sector.
The DSB sector is based on a SUSY SU(2) gauge theory with four doublet
chiral superfields $Q_i$ where $i$ is a flavor index ($i=1,\cdots,4$).
Here, we have a global flavor SU(4)$_F$.
We assume the following tree-level superpotential introducing six
singlet chiral superfields $Z$ and $Z^a$ ($a=1,\cdots,5$):
\begin{eqnarray}
  W_{\rm tree} = \lambda Z (QQ) + \lambda_Z Z^a (QQ)_a,
\label{tree-sup}
\end{eqnarray}
where $(QQ)$ and $Z$ are singlets of the SP(4)$_F$ subgroup of the
flavor SU(4)$_F$ and $(QQ)_a$ and $Z^a$ are five-dimensional
representations of the SP(4)$_F$.
As shown in Ref.~\cite{HINTY}, integration of the SU(2) gauge fields
together with $Q_i$ and $Z^a$ leads to the low-energy effective
superpotential
\begin{eqnarray}
  W_{\rm eff} \simeq \frac{\lambda}{(4\pi)^2} \Lambda^2 Z,
\end{eqnarray}
for $\lambda_Z > \lambda$, where $\Lambda$ is a dynamical scale of the
SU(2) gauge interaction.\footnote{
The factor of $4\pi$ is determined by the na\"{\i}ve dimensional
analysis \cite{NDA}.}
We have nonvanishing $F$ term, $\vev{F_Z} \simeq \lambda \Lambda^2 /
(4\pi)^2 \neq 0$, and hence SUSY is broken \cite{IYIT}.
We also assume that the fields in the DSB sector are charged under the
U(1)$_m$ gauge interaction.
The charge assignments of chiral superfields are given by \cite{NTY}
\begin{equation}
  Q_{1} (+1), \quad Q_{2} (-1), \quad Q_{3} (0), \quad Q_{4} (0).
\end{equation}
Here, the numbers in each parentheses denote the U(1)$_m$ charges.
The U(1)$_m$ charges for $Z$ and $Z^a$ are determined such that the
superpotential Eq.~(\ref{tree-sup}) is invariant under the U(1)$_m$.

We now turn to the messenger sector.
It consists of three chiral superfields,
\begin{eqnarray}
  E (+1), \quad \bar{E} (-1), \quad S (0),
\end{eqnarray}
and vector-like messenger quark and lepton superfields, 
$d$, $\bar{d}$, $l$, $\bar{l}$.
Here, the messenger quark multiplets $d$, $\bar{d}$ and lepton
multiplets $l$, $\bar{l}$ are all neutral under the U(1)$_m$.
The $d$ and $\bar{d}$ ($l$ and $\bar{l}$) transform as the right-handed
down quark and its antiparticle (the left-handed lepton doublet and its
antiparticle) under the standard-model gauge group, respectively.
The superpotential for the messenger sector is given by
\begin{eqnarray}
  W_{\rm mess} = k_E S E \bar{E} + \frac{f}{3} S^3 
    + k_d S d \bar{d} + k_l S l \bar{l}.
\label{superpotential}
\end{eqnarray}
The SUSY-breaking effects in the DSB sector are transmitted to the
messenger sector through the U(1)$_m$ gauge interaction.
As a result, the $E$ and $\bar{E}$ fields obtain positive soft
SUSY-breaking squared masses $m_E^2$ and $m_{\bar{E}}^2$,\footnote{
If $\vev{Z}=0$, there is an unbroken U(1) $R$-symmetry.
In this case, the U(1)$_m$ gaugino remains massless while $E$ and
$\bar{E}$ fields obtain soft SUSY-breaking masses in
Eq.~(\ref{mass_E}).}
\begin{eqnarray}
  m_E = m_{\bar{E}} 
    \sim \frac{\alpha_m}{4\pi}\frac{\lambda F_Z}{\Lambda} 
    \simeq \frac{\alpha_m}{4\pi}\frac{\lambda^2}{16\pi^2}\Lambda,
\label{mass_E}
\end{eqnarray}
where $\alpha_m = g_m^2/4\pi$ is the U(1)$_m$ gauge coupling constant.
They generate the following negative soft SUSY-breaking mass squared
$-m_S^2$ for the $S$ field through the Yukawa coupling $k_E S E \bar{E}$ 
in Eq.~(\ref{superpotential}) at the one-loop level:
\begin{eqnarray}
  m_S^2 \simeq \frac{4}{(4\pi)^2}k_E^2 m_E^2 \ln\frac{\Lambda}{m_E}. 
\end{eqnarray}

Altogether, the resulting scalar potential is 
\begin{eqnarray}
  V_{\rm mess} = \sum_\eta |\frac{\partial W_{\rm mess}}{\partial \eta}|^2 
    + m_E^2 |E|^2 + m_{\bar{E}}^2 |\bar{E}|^2 - m_S^2 |S|^2,
\label{scalar_potential}
\end{eqnarray}
where $\eta$ denotes chiral superfields $E, \bar{E}, S, d, \bar{d}, l$
and $\bar{l}$.
This potential has a global minimum at 
\begin{eqnarray}
  && \vev{S^* S} = \frac{m_S^2}{2 f^2}, \qquad
     \vev{|F_S|} = \frac{m_S^2}{2 f}, \nonumber\\
  && \vev{E} = \vev{\bar{E}} = \vev{d} = \vev{\bar{d}} =
     \vev{l} = \vev{\bar{l}} = 0,
\end{eqnarray}
in a certain parameter region \cite{NTY}.
Thus, all the standard-model gauge symmetries are preserved and the
SUSY-breaking effects are transmitted to the messenger quark and lepton
multiplets through $\vev{F_S}$.

The soft SUSY-breaking masses for the gauginos $\tilde{g}_i$
($i=1,\cdots,3$) and the squarks, sleptons, and Higgses $\tilde{f}$ in
the MSSM sector are generated by integrating out the messenger quarks
and leptons as
\begin{eqnarray}
  m_{\tilde{g}_i} &=& c_i \frac{\alpha_i}{4\pi} \Lambda_{\rm mess}, \\
  m_{\tilde{f}}^2 &=& 2 \Lambda_{\rm mess}^2 \left[ 
    C_3 (\frac{\alpha_3}{4\pi})^2 + 
    C_2 (\frac{\alpha_2}{4\pi})^2 + 
    \frac53 Y^2 (\frac{\alpha_1}{4\pi})^2  \right], 
\end{eqnarray}
where $c_1 = 5/3$, $c_2 = c_3 = 1$; $C_3 = 4/3$ for color triplets and
zero for singlets, $C_2 = 3/4$ for weak doublets and zero for singlets,
and $Y$ is the hypercharge ($Y = Q_{\rm em} - T_3$).
Here, $\Lambda_{\rm mess}$ is an effective messenger scale defined as 
\begin{eqnarray}
  \Lambda_{\rm mess} \equiv \frac{\vev{|F_S|}}{\vev{|S|}} 
    = \frac{m_S}{\sqrt{2}}, 
\end{eqnarray}
which can be written in terms of the SUSY-breaking scale $\sqrt{F_Z}$ as 
\begin{eqnarray}
  \Lambda_{\rm mess} &\simeq& \frac{\sqrt{2}}{(4\pi)^4} \alpha_m 
    \lambda^2 k_E \sqrt{\ln \frac{(4\pi)^3}{\alpha_m \lambda^2}} 
    \cdot \Lambda \nonumber\\
  &=& \frac{\sqrt{2}}{(4\pi)^3} \alpha_m \lambda \sqrt{\lambda} k_E 
    \sqrt{\ln \frac{(4\pi)^3}{\alpha_m \lambda^2}} 
    \, \sqrt{F_Z}.
\label{L_mess}
\end{eqnarray}

We are now at the point of this letter.
We consider that the above model is the low-energy effective theory of
the brane world, in which all fields in the DSB sector reside on the
hidden brane while the messenger and MSSM sectors are localized on the
observable brane.
Then, the U(1)$_m$ gauge multiplet should necessarily live in the bulk.
Thus, we identify U(1)$_m$ with the bulk U(1)$_{\rm bulk}$.\footnote{
We assume that the compactification scale $L^{-1}$ is sufficiently high
as $L^{-1} \gg \Lambda$.}
The SUSY breaking on the hidden brane is transmitted to the observable
brane by the U(1)$_{\rm bulk}$ gauge and gravitational interactions
across the bulk between two branes.

The effective messenger scale $\Lambda_{\rm mess}$ should be taken at
$(10^4-10^5)~\GEV$ to induce the MSSM gaugino and sfermion masses of the 
electroweak scale.
On the other hand, the supergravity effects generate simultaneously the
$\mu$ term and the gravitino mass as discussed in the introduction
\cite{GM},
\begin{eqnarray}
  \mu \simeq m_{3/2} = \frac{F_Z}{\sqrt{3}M_G}.
\end{eqnarray}
We should set $\sqrt{F_Z} \simeq (2-6) \times 10^{10}~\GEV$ to reproduce 
correctly the electroweak symmetry breaking ({\it i.e.} $\mu \simeq
100~\GEV-1~\TEV$).
It is a crucial point that the above two conditions determine the 
U(1)$_{\rm bulk}$ gauge coupling constant through Eq.~(\ref{L_mess}).
Assuming the Yukawa coupling constants $\lambda$ and $k_E$ connecting
fields on the same brane to be of order one, we obtain the 
U(1)$_{\rm bulk}$ gauge coupling 
$\alpha_{\rm bulk} \simeq 5 \times 10^{-4}$ at the scale $\Lambda$.
Since the running effect of the U(1)$_{\rm bulk}$ gauge coupling is
negligible for the matter content in the present model, we find that 
$\alpha_{\rm bulk} \simeq 5 \times 10^{-4}$ at the compactification
scale of the extra dimension.

For a definiteness, we here assume the $(4+1)$-dimensional spacetime
with one extra dimension compactified on the orbifold $S^1/{\bf Z}_2$.
The hidden and observable branes are located at two different fixed
points of the $S^1/{\bf Z}_2$ separated by a distance $L$.
In this case, the 5-dimensional U(1)$_{\rm bulk}$ multiplet is composed
of a vector field, a Dirac spinor field and a real scalar field, which
corresponds to a ${\cal N} = 2$ vector multiplet in 4 dimensions.
Through an orbifold projection, however, only the 4-dimensional 
${\cal N} = 1$ vector multiplet of the U(1)$_{\rm bulk}$ can couple to
the fields on two branes.\footnote{
A detained analysis on the orbifold $S^1/{\bf Z}_2$ is given in
Ref.~\cite{MP}.}

The 4-dimensional gauge coupling $g_{\rm bulk}$ is obtained from the
5-dimensional coupling $g_{\rm bulk}^{(5)}$ as
\begin{eqnarray}
  \frac{1}{g_{\rm bulk}^2} = \frac{2L}{(g_{\rm bulk}^{(5)})^2}.
\end{eqnarray}
Assuming that $g_{\rm bulk}^{(5)}$ is of order one in the unit of the
fundamental scale $M_*$, we obtain the following relation:
\begin{eqnarray}
  (2L)^{-1} = (4\pi\alpha_{\rm bulk}) M_*,
\label{relation-1}
\end{eqnarray}
which shows that the compactification length $L$ is 
$(8\pi\alpha_{\rm bulk})^{-1} \simeq 100$ times larger than the
fundamental length scale $M_*^{-1}$ of the theory.
This sufficiently suppresses the unwanted FCNC's which would be induced
by exchanges of bulk fields of masses around $M_*$ \cite{RS}.

On the other hand, the gravitational scale $M_G$ is given by
\begin{eqnarray}
  M_G^2 = 2 M_*^3 L.
\label{relation-2}
\end{eqnarray}
Thus, together with Eq.~(\ref{relation-1}), we find that the fundamental 
scale $M_*$ and the compactification scale $L^{-1}$ are given by
\begin{eqnarray}
  && M_* = M_G (4\pi\alpha_{\rm bulk})^{1/2} \simeq 2 \times 10^{17}~\GEV, \\
  && L^{-1} = 2M_G (4\pi\alpha_{\rm bulk})^{3/2} \simeq 2 \times 10^{15}~\GEV.
\end{eqnarray}
It is very interesting that these values are close to the ones discussed
in Ref.~\cite{bulk}.

Several comments are in order.
First, the U(1)$_{\rm bulk}$ gauge symmetry is unbroken in the present
model so that one of scalar components of the $E$ and $\bar{E}$ fields
is completely stable.
This requires the reheating temperature $T_R$ of inflation to be lower
than the mass of the lightest scalar field of order 
$\vev{S} \simeq 10^5~\GEV$, in order for its present energy density not
to exceed the critical density of the universe.
Second, the dangerous $D$-term for the U(1)$_{\rm bulk}$ does not appear
in the model \cite{NTY}, since there is an unbroken charge conjugation
symmetry defined as 
\begin{equation}
\begin{array}{c}
  Q_1 \rightarrow  Q_2, \quad
  Q_2 \rightarrow -Q_1, \nonumber\\
\\
  V_{\rm bulk} \rightarrow -V_{\rm bulk}, \quad
  E \rightarrow \bar{E}, \quad
  \bar{E} \rightarrow E,
\end{array}
\label{cc-sym}
\end{equation}
where $V_{\rm bulk}$ is the U(1)$_{\rm bulk}$ gauge superfield.
The singlets $Z$ and $Z^a$ are assumed to transform properly so that the 
superpotential Eq.~(\ref{tree-sup}) is invariant under the charge
conjugation symmetry Eq.~(\ref{cc-sym}).
This is a consequence of the vector-like structure of the 
U(1)$_{\rm bulk}$ gauge sector.
Third, if there exists a non-Abelian gauge theory in the bulk other
than the U(1)$_{\rm bulk}$, the radius of the extra dimension is
stabilized as shown in Ref.~\cite{LS}.\footnote{
The radius can also be stabilized by the mechanism of Ref.~\cite{IY}, if 
there are two non-Abelian gauge theories in the bulk which couple to
suitable matters on a brane.}
Note that the stabilization is not disrupted by the Casimir energy
induced by the SUSY breaking \cite{MP}.

Finally, we should stress that the gravitino has a mass of order
$100~\GEV-1~\TEV$ in the present model although the mass spectrum of the
other SUSY particles is the same as that of the gauge-mediated SUSY
breaking models.\footnote{
The $B$ term is of the order of the gravitino mass, where $B$ is defined
as ${\cal L} = \mu B \tilde{H} \tilde{\bar{H}} + {\rm h.c.}$
($\tilde{H}$ and $\tilde{\bar{H}}$ are the scalar components of $H$ and
$\bar{H}$).
Thus, the small $\tan\beta$ region may be accommodated in the present
model.}
This leads to an interesting phenomenological consequence.
That is, the bino is most likely the dark matter in the present
universe while the usual gauge mediation models predict the gravitino
dark matter.
It is a generic feature \cite{INY} of the gauge-mediated SUSY breaking
models in which the phenomenologically viable $\mu$ term arises from the
supergravity effect \cite{GM}.
We should stress that the Randall-Sundrum brane-world scenario \cite{RS} 
provides a natural solution to the $\mu$ problem in a large class of
gauge-mediated SUSY breaking models \cite{min-GMSB, NTY}, although we
have adopted a specific model in Ref.~\cite{NTY} to demonstrate our
point.

\vspace{7mm}

{\bf Acknowledgments}

We would like to thank Izawa K.-I. for a useful discussion.
Y.N. thanks the Japan Society for the Promotion of Science for financial 
support.
This work was partially supported by ``Priority Area: Supersymmetry and
Unified Theory of Elementary Particles (No. 707)'' (T.Y.).

\newpage
%
%
%
\newcommand{\Journal}[4]{{\sl #1} {\bf #2} {(#3)} {#4}}
\newcommand{\PL}{\sl Phys. Lett.}
\newcommand{\PR}{\sl Phys. Rev.}
\newcommand{\PRL}{\sl Phys. Rev. Lett.}
\newcommand{\NP}{\sl Nucl. Phys.}
\newcommand{\ZP}{\sl Z. Phys.}
\newcommand{\PTP}{\sl Prog. Theor. Phys.}
\newcommand{\NC}{\sl Nuovo Cimento}
\newcommand{\MPL}{\sl Mod. Phys. Lett.}
\newcommand{\PRep}{\sl Phys. Rept.}

%
\end{document}